\documentstyle[multicol,aps,prl,epsf]{revtex}
\def\ba{\begin{array}}
\def\ea{\end{array}}
\def\be{\begin{equation}\begin{array}{l}}
\def\ee{\end{array}\end{equation}}
\def\bea{\begin{equation}\begin{array}{l}}
\def\eea{\end{array}\end{equation}}
\def\f#1#2{\frac{\displaystyle #1}{\displaystyle #2}}
\def\om{\omega}

\def\de{\delta}
\def\De{\Delta}
\def\va{\varepsilon}

\def\la{\lambda}

\def\bi{\bibitem}
\def\c{\cite}

\def\lan{\langle}
\def\ra{\rangle}

\title{Delocalization of Wannier-Stark ladders by 
phonons: tunneling and stretched polarons}

\author{Wei Zhang,$^{a}$ Alexander O. Govorov,$^{a,b}$ and Sergio E. Ulloa$^{a}$\\
$^{a}$Department of Physics and Astronomy, and Condensed Matter
and Surface Sciences Program, Ohio University, Athens, Ohio
45701--2979 \\
 $^b$Institute of Semiconductor Physics, RAS,
Siberian Branch, 630090 Novosibirsk, Russia }

\begin{document}

\maketitle

\begin{abstract}
We study the coherent dynamics of a Holstein polaron in strong
electric fields. A detailed analytical and numerical analysis
shows that even for small hopping constant and weak
electron-phonon interaction, polaron states can become delocalized
if a resonance condition develops between the original
Wannier-Stark states and the phonon modes, yielding both tunneling
and `stretched' polarons.  The unusual stretched polarons are
characterized by a phonon cloud that {\em trails} the electron,
instead of accompanying it.  In general, our novel approach allows
us to show that the polaron spectrum has a complex nearly-fractal
structure, due to the coherent coupling between states in the
Cayley tree which describes the relevant Hilbert space.  The
eigenstates of a finite ladder are analyzed in terms of the
observable tunneling and optical properties of the system.

\end{abstract}
The coupling of an electronic system with a coherent boson field
has been a subject of great interest in recent years, thanks to
the availability of high quality materials and intense photon
sources.  In condensed matter systems,
however, a natural intrinsic field of bosonic nature is provided
by the lattice vibrations.  The coupling to phonons is typically
considered to result in inelastic {\em incoherent} scattering for
the electron, as the phonons are difficult to interrogate
separately. This results in an electronic `open' system which
exists immersed in the assumed-incoherent phonon field.  Although
this description is appropriate for many situations, either
because the electron-phonon coupling is weak and/or non-resonant,
or if the system is at high temperatures, its validity is suspect
if the interaction is effectively strong and the temperature is
low.

Electronic transport in superlattices has yielded a number of
interesting phenomena and concepts, including Bloch oscillations
and Wannier-Stark ({\bf WS}) ladders \c{exp1}, as well as negative
differential conductance \c{con} in high DC electric fields,
dynamical localization \c{dy}, fractional WS ladders under DC and
AC electric fields \c{fws}, and resonant magnetopolarons \c{mpl}.
Similarly, electron-phonon interactions under high electric fields
have been investigated in some detail, including coupled
Bloch-phonon and -plasmon oscillations \c{blo}, as well as
phonon-assisted--hopping of an electron on a WS ladder \c{pho1}. A
powerful variational treatment of inelastic but coherent quantum
transport was presented in \c{tru}, while anomalies in transport
properties under a resonant condition were studied in \c{res}, and
the optical absorption associated with the resonance of a WS
ladder and confined optical phonons was studied in Refs.\
\c{gov1,gov2}.

In this paper, we study the effects of coherent coupling of an
electron to the phonon system present in its solid at low
temperatures. We are specially interested in the effect of this
coupling on the tunneling properties of an electron in an intense
electric field region, like the situation achieved in
semiconductor superlattices, for example \cite{sys-note}. Our
description is non-perturbative for the resonant case with
exponentially large Hilbert space, and as such we are able to
elucidate the role of resonant phonon fields on the otherwise
localized electron residing in a WS ladder, for both weak and
strong coupling. Using a Holstein model for the electron-phonon
coupling, and both analytical and numerical techniques, we
demonstrate that the electron tends to become more extended for
increasing coupling with phonons. Moreover, we show that resonant
coupling results in {\em total delocalization} of the polaron for
some of the states, and the consequent restoring of a `miniband'
structure in the system, even as the applied electric field
remains strong. Other `stretched polaron' states, however, are
highly degenerate yet strongly localized on one end of a finite
structure, with a phonon `cloud' which is nearly detached from the
electron (somewhat a precursor of the polaron dissociation
described in \c{Conwell}). We are able to show explicitly that the
level spectrum in this regime has a nearly fractal structure with
rather complex wave functions. On the other hand, when the system
is away from the resonant condition, a deformed WS ladder results,
with electronic wave functions which nevertheless remain
localized. We finally analyze the tunneling and optical properties
one could observe experimentally in such a system.

We consider a Holstein model \c{holst}, describing a single
electron in a one-dimensional tight-binding lattice which
interacts locally with dispersionless optical phonons.  In
addition, we consider that a strong constant electric field is
applied which generates a WS ladder. The Hamiltonian is then
 \be
 H_0=\sum_j \left\{ \va_j c^+_jc_j+t (c_j^+c_{j+1}+c^+_{j+1}c_j) \right\} \\
[1ex] ~~~~~ + \om \sum_j a_j^+a_j+\gamma \sum_j c_j^+c_j(a_j^+
+a_j),
 \ee
 where $t$ is the electron hopping constant, $\om$ the phonon
frequency, $\gamma$ the electron-phonon coupling constant,
$\va_j=-edEj \equiv -j \Delta$ the site (or `well') energy, $E$
the electric field, $d$ the lattice constant, and $\Delta$ the
spacing between WS `rungs' (we have set $\hbar=1$). It is known
that well-localized WS states will form in the absence of
 electron-phonon interactions, with eigenvalues and
eigenfunctions given by $\va_j$, and $| \phi_j \rangle =\sum_i
J_{i-j} (2t/edE) \, |i\rangle $, where $J_{l}$ is the $l$-th order
Bessel function \cite{dy}. 
 Given the properties of $J_l$, the WS
$j$-state is clearly localized with a characteristic length
$2t/eE$ around site $j$. It is helpful to introduce the WS
creation and annihilation operators, $ 
  d_j=\sum_{i=-\infty}^{\infty} J_{i-j}(2t/edE) c_i \, .
$ 
 It is easy to show that $\{d^+_i,d_j\}=\de_{ij}$, and that the
Hamiltonian can be rewritten in terms of $d_j,d^+_j$. In the case
of strong electric field (or small hopping constant) ($2t/\De \ll
1$), the Hamiltonian can
be simplified to
 \bea H = \sum_j \left \{- \De j
d_j^+d_j+ \om  a_j^+a_j +\gamma d_j^+ d_j (a_j^+ +a_j) \right. \\
 [1ex] ~~~~~~ - \left. \la (a_j+a_j^+ -a_{j+1}^+-a_{j+1})
(d_j^+d_{j+1}+d_{j+1}^+ d_j) \right\} \, ,
 \label{old7}
 \eea where the effective hopping becomes $\la=\gamma t/\De$.
Notice that phonon-assisted hopping develops between the WS states
via the last term.
Eq. (\ref{old7}) is similar to that in \c{gov1},
but here, as the electron jumps
it can not only emit (or absorb) a phonon on its current location,
but also on the neighboring site. This  more
natural description results in a much larger
relevant Hilbert space for this problem.

Consider the process of an electron jumping between WS states and
creating or annihilating a phonon with amplitude $\lambda$, as per
the last term in (\ref{old7}). The relevant Hilbert space can be
described for $n=3$ via the basis: $|1;000\rangle, |2;010\rangle,
|2;100\rangle, |3;011\rangle, |3;101\rangle, |3;020\rangle$, and
$|3;110\rangle$. Each state $|j;m_1,m_2,... \rangle$ is indexed by
the electron position $j$, and $m_k$ refers to the number of
phonons on site $k$.  This construction yields a Hilbert space
of dimension $N-1=2^{n} -1$, where $n$ is the total number of
sites in the chain. It is interesting to note that the structure
of this vector space is that of a Cayley tree where each $j$-state
links to two at $j+1$, in general with non-symmetric weights in
all branches \c{car}. A similar structure of the vector space was
found in \c{tru}.

Under the resonance condition, i.e., $\De= \om$, all the
basis-states in the Hilbert space described above have the {\em
same energy}, and one would expect that the off-diagonal matrix
elements in (\ref{old7}) would break the degeneracy. One would
expect that a `miniband' of extended states might form, whenever
the resonance condition is reached.
For this case,
Hamiltonian (\ref{old7}) in the above basis of $n=3$ takes the form
 \small \bea
 H=\left( \ba{ccccccc}
-\De & \la & -\la & 0 & 0 & 0 & 0\\
\la & -\De & 0  &\la & 0 & -\sqrt{2}\la & 0 \\
-\la & 0 & -\De & 0 & \la & 0 & -\la \\
0 & \la & 0 & -\De & 0 & 0 & 0\\
0 & 0 & \la & 0 & -\De & 0 & 0\\
0 & -\sqrt{2} \la & 0 & 0 & 0 & -\De & 0\\
0 & 0 & -\la & 0 & 0 & 0 & -\De \ea \right) \, ,
 \eea \normalsize
 with ever larger blocks for larger $n$ values.  Notice that there
are elements with somewhat unusual values, $-\sqrt{2}\la$,
associated with the higher number of phonons in state
$|3;020\rangle$.  A similar structure is seen for all values of
$n$, and as these terms appear only sporadically, they have only a
small influence on the overall level structure.  In particular,
one expects that there would be little change in the physics in
the limit of large $N$, if we replace $-\sqrt{2}\la $ terms with
$-\la$. This assumption is in fact confirmed explicitly by
numerical calculations. Moreover, this `symmetrized' Hamiltonian
$H_{sym}$ allows one to better analyze the structure of the
solutions of $H$, as we show below, while exhibiting some of the
same generic features and fully self-similar properties.

{\em Symmetrized Hamiltonian}. By using the block decomposition of
the determinant, 
$ \det \left( \ba{cc}
A & B\\
C & D
 \ea \right) = \det(D) \det (A-BD^{-1}C) \, ,
$ 
 one can find that the eigenvalues of $H_{sym}$ are determined by
the equation $ 
 \va_0^{N/2} \va_1^{N/4}
\va_2^{N/8} \cdots \va_{n-1}^1=0 \, ,
 \label{symeigen}
 $ 
 where $\va_0=\va$, $\va$ is the energy eigenvalue,
  and  $ 
\va_{k+1}=\va- 2\la^2/ \va_k \, . $ 
Here we have made a constant energy shift of $\De$.
One can then write $\va_k$ as a continuous fraction in $k$ steps
 \be
\va_k=\va-\f{2\la^2}{\va-\f{2\la^2}{\va  -
\f{.}{...}}} \, .
 \ee

and the eigenvalues of $H_{sym}$ are
given by \c{gov2} \c{frac}
 \be \va_{k,l}=2\sqrt{2} \la
 \cos \left(\f{l\pi}{k+1} \right) \, , \label{band-sym}
 \ee
where $k=1,2,...,n$; $l=1,...,k$. The limiting bandwidth of the
spectrum is $4 \sqrt{2} \la$.

Using $H_{sym}$, one can arrive at an estimate for the degeneracies
for the first few
eigenvalues in the limit of large $N$, given by the equation
$\va^{3N/8} (\va^2-2\la^2)^{N/8} (\va^2-4\la^2)^{N/8}=0$.  
One then has eigenvalues $\va=0$ with degeneracy $3N/8$,
$\va=\pm\sqrt{2}\la$ with degeneracy $N/8$, and $\va=\pm 2 \la$
with degeneracy $N/8$, in agreement with our explicit calculations
for finite $N$.  Notice that increasing lattice or chain size
produces both new energy eigenvalues within the limiting band and
additional states which augment the degeneracy of some of the old
eigenvalues.The overall structure of the spectrum remains the same and
all the eigenvalues of the shorter chain are also in the spectrum of the
longer one \c{more-details}.

Figure 1 shows typical numerical results for the eigenvalues of
the physical Hamiltonian $H$ in Eq.\ (\ref{old7}), while the inset
shows the results for the symmetrized version $H_{sym}$ (both for
$n=8$). Our analytical results match exactly those shown in the
inset, which in turn mimic well the results for $H$ in the main
panel, although small differences appear due to the sporadic
$-\sqrt{2} \lambda$ terms. The self-similar structure of the
spectrum in the inset is the manifestation of the full
self-similarity of the Hamiltonian $H_{sym}$ (while it is only
approximate for $H$). One can see that the original highly
degenerate Hilbert basis is split into a semi-continuous `miniband' by
the off-diagonal elements in $H$ (and which for $H_{sym}$ yields
Eq.\ \ref{band-sym} for large $N$). Notice, however, that there
are large residual degeneracies for the eigenvalues of $H$ in the
center of the miniband. The right panel in Fig.\ 1 shows the
density of states of the $H$ system, exhibiting also an
approximate self-similarity.

The `electronic wave functions' for various sample eigenstates are
shown in Fig.\ 2, $P_\va (j)= \sum_{\{m\}} |C^\va_{j,\{m\}}|^2$,
where the $\va$-eigenstate is given in terms of the basis above by
$|\Psi_\va \rangle = \sum_{j,\{m\}} C^\va_{j,\{m\}} |j;\{m\}
\rangle$. One can observe that the states in regions of high
degeneracy, $\va =0$ and $\va/\la \approx -1.6$ (where states (A) and
(C) are indicated), are localized around the right end of the
structure, $j\approx 8$. However, non-degenerate states in the
band, such as the one labeled (B), result to be fully extended
over the structure.

Notice that in this inherently coupled electron-phonon system, the
calculated eigenstates are a coherent mixture of electron and
phonons on different sites.  The amplitudes in Fig.\ 2 have been
projected over all phonon components. Similarly, in order to
illustrate the physics of these states better, one can exhibit the
average phonon number for various states in Fig.\ 3. One can see
clearly that all states at the degenerate center of the band have
many more phonons ($\simeq 7$) than those non-degenerate states
residing higher in the band.  It appears, in fact, that when the
electron interacts with many phonons, it localizes, as one would
perhaps intuitively expect for polaron self-trapping states, while
states with fewer phonons are less localized and `propagate'
better. One can provide an estimate for the average number of
phonons for extended states by the following argument: The average
phonon number is given by $\langle N \rangle=\sum_i P_i N_i$,
where $P_i$ is the probability for an electron to be at site $i$,
and $N_i$ is the number of phonons in this state. For an extended
state, $P_i \simeq 1/n$, where $n$ is the number of sites. A state
with an electron at site $i$ can be obtained by the electron
hopping $i-1$ steps from the first site, while emitting $N_i=i-1$
phonons in the process.  We then obtain that the average number of
phonons for an extended states is $\langle N \rangle \simeq \sum_i
(i-1)/n=(n-1)/2$. This result matches well the numerical results,
where the most extended states have $\langle N \rangle \simeq 3.5$
in Fig.\ 3, such as the state labeled (B).

One can also inspect the spatial distribution of phonons for each
state, as in the inset of Fig.\ 2, which shows the corresponding
phonon distribution for several eigenstates, $N_\va (j) =
\sum_{l,\{m_j\}} |C^\va_{l,\{m_j\}}|^2 m_j$, indicating how the
phonons distribute over the chain.  The degenerate states, such as
(A) and (C), with the electron localized to the right of the
structure, have nearly one phonon per site throughout the chain,
indicating that these polarons have a `stretched' phonon cloud
away from the charge, seemingly in anticipation of the total
polaron dissociation that other authors have described for
extended chains under high electric fields \cite{Conwell}.  In
contrast, the most delocalized states (such as (B) in Fig.\ 2)
show a phonon amplitude much less than unity throughout, yielding
an overall low phonon count, and a cloud that in general
`accompanies' the charge.

{\em Interband optical absorption}. The spectrum of optical
absorption associated with transitions between valence and
conduction bands provides information about the density of states
and the states' charge distribution.
The absorption intensity $K$ is calculated from
$K=\f{2\pi}{\hbar}|W_\nu|^2\de(\va_\nu+E_g-\om)$,
where $E_g$ is the valence and conduction band gap,
$\nu$ refers to the various
polaron states, and $W_\nu$ is the transition matrix between valence
band and conduction band states. Under the strong field present in the system,
the hole is localized the most because of its large effective mass. Thus we
ignore its negligible tunneling and coupling to phonons and assume it
localized at site 1. Correspondingly, there is
 conservation of phonon number and  $|W_\nu|^2 \propto P_\nu(1)$ \c{note2}.
In Fig.\  4(a) we show the absorption intensity vs.\ phonon
frequency.  Once again, we see that the non-degenerate states,
$\va/\la \simeq \pm 0.6$, have the largest amplitude at site 1 and
will then contribute the most to the absorption spectrum.  In
contrast, {\em all} highly degenerate states at $\va = 0$ are
strongly localized at the right end of the chain and make a
vanishing contribution to $K$. Such strong modulation of $K$ would
be easily accessible to experiments.

{\em Transport properties}.  As described above, tunneling
experiments under strong electric fields are possible in
semiconductor superlattices and other systems, and it is of
interest to investigate the role of the resonance condition.
The transition probability $P$ can be calculated in a $S$-matrix
formalism. $P=|T|^2$ and $T=\lan\va_f,R|S|\va_i,L\ra$. Here $R$ and
$L$ refer to the right and left leads. In the wide band limit,
  $T\propto \Gamma \int dt_1 dt_2e^{i(\va_f t_2-\va_i t_1)}
  \lan n|G^{R}(t_2-t_1)|1\ra
  =\sum_\nu \Gamma \lan n|\nu\ra\lan\nu|1\ra\de(\va_f-\va_i)$,
  where
 $\Gamma$ describes the electron interaction with contacts, and $G^{R}$ is
 the retarded Green function connecting both ends of the structure
 (the site $j=1$ and $j=n$)  \c{tul}.
 We use $D(\nu)=|\lan n|\nu\ra\lan\nu|1\ra|^2
 =P_\nu(1)P_\nu(n)$ to describe the relative contribution of various
 eigenstates $\nu$ to the electronic transport, as
the tunneling probability is proportional to the density of states in
the leads and  $D(\nu)$
\cite{tul}.
 Figure 4(b)
shows the quantity $D(\va)$. We can see that {\em all} the
states at the center of the band contribute {\em zero} to the
transport amplitude through the chain, and that this behavior is
exhibited by all the high peaks in the DOS, clearly consistent
with their spatially localized-charge nature. On the other hand,
$D$ shows large values for extended states, confirming in fact
that the non-degenerate states in the miniband are extended.
The variations shown in $D$ would then be reflected in
strong amplitude modulations within each of the phonon replicas in
tunneling experiments
\cite{exp2},
whenever the resonance condition is reached.
It is also clear that despite the fact that $K$ and $D$ correspond
to completely different physical quantities, their energy behavior
is surprisingly similar.  This arises because of the large
electric field asymmetry, resulting in no states with mostly/only
large electronic amplitude in the first site (i.e., localized
towards the left of the chain).
Once a state is extended and has
non-vanishing amplitudes at site 1, it also does at site $n$.

So far we have mainly described the resonant case.
When the system is away from the resonant condition, a deformed
lattice with substructure is formed. We are also able to study
the system beyond small hopping and weak coupling regime by mapping
it via the Lang-Firsov canonical transformation to the case we have
studied, with renormalized hopping
and coupling constants. In this approach, we can provide additional
physical insights into some interesting results found in \c{tru}.
We will report the details elsewhere
\c{more-details}.

Although our model is 1D \c{sys-note}, we expect the qualitative
features and main conclusions to be valid for semiconductor superlattices (SLs)
where electron--optical-phonon coupling is highly anisotropic.
In that case, it would be possible to neglect the {\em in-plane} scattering
of electrons by phonons, while the strong interaction assists in the electron
hopping. In fact, for SLs with low in-plane disorder, the transitions between different
layers involve states near the bottom of the 2D electron subbands and
phonons with small in-plane momenta. Hence, the system would exhibit
quasi-1D behavior, and in-plane scatterings are a small correction.
Similarly, we could adapt our model to the description of polarons in
polymer chains, such as those described in Ref. \c{Conwell}, although
there the electron-phonon interaction is of different strength than in
solid SLs, and tunneling experiments are rather more challenging.

We have studied the coherent dynamics of Holstein polarons in a
strong electric field. It is found that with the help of phonons,
each WS rung develops into a `miniband' under the resonance
condition. This miniband shows a nearly self-similar structure,
which is inherited from the full self-similarity of the
Hamiltonian $H_{sym}$ for a symmetric Cayley tree.  Although the
phonons can help the electron jump from one WS state to another,
the phonons can also {\em prevent} the electron from propagating.
If too many phonons are involved, this results in highly
degenerate states with the electron localized at one end of the
structure while the phonons are pulled away from it (the
`stretched' polaron). The miniband structure and its modulations
in density of states and other characteristics are manifested in
both transport and optical properties of the system, which we
anticipate can be observed in experiments at low temperatures
\cite{sys-note}.

\acknowledgments
This work was supported in part by US DOE No.\
DE--FG02--91ER45334.

\newpage

\begin{figure}
\centerline{ \epsfxsize=3.3in \epsfbox{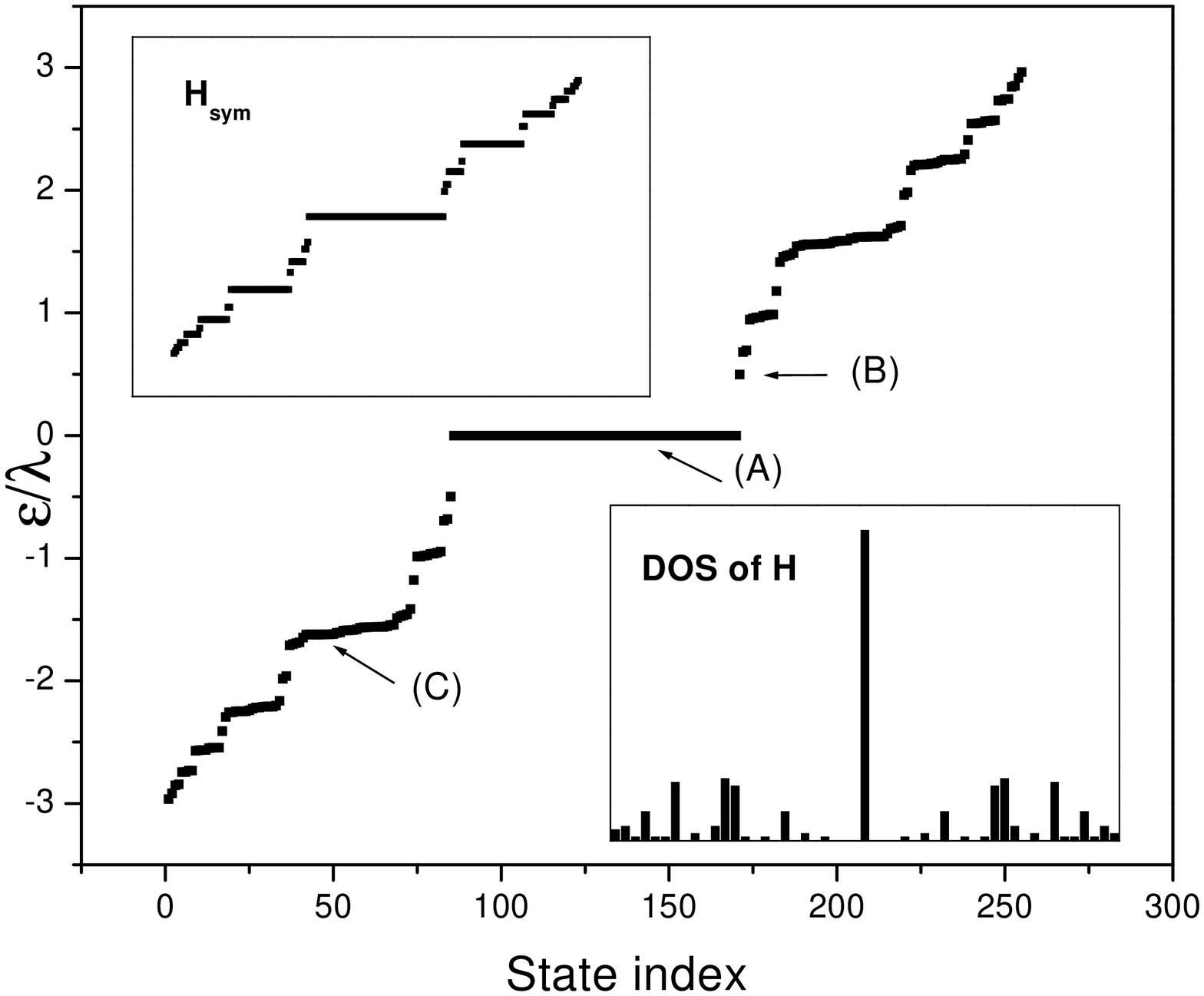}}
\caption{Energy spectrum for $H$ (main fig.) and $H_{sym}$ (left
inset) for a system with lattice size $n=8$. Notice clear fractal
structure of spectrum for $H_{sym}$, and only partially for $H$.
Right panel shows DOS of $H$, fully symmetric about $\varepsilon
=0$. Labels A to C relate to Fig.\ 2. Here,
$\De=\om=10\la .$}
 \label{fig1}
\end{figure}

\begin{figure}
\centerline{ \epsfxsize=3.1in \epsfbox{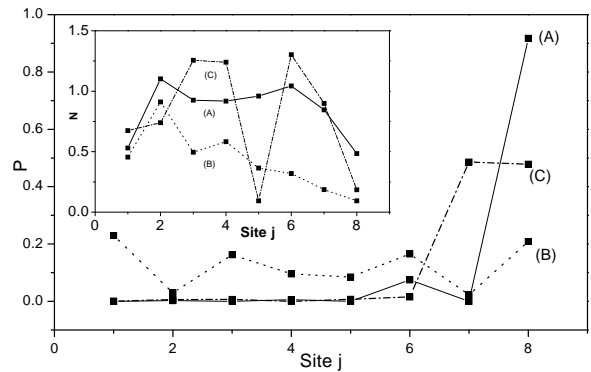}}
\caption{Electronic amplitudes $P(j)$ at each site $j$ for several
states, A to C, as indicated in Fig.\ 1. Notice A and C are
localized near $j \simeq 8$.  Panel shows phonon content $N(j)$
for same states.  A and C have large phonon counts throughout and
away from the electron, resulting in `stretched' polarons. }
 \label{fig2}
\end{figure}

\begin{figure}
\centerline{\epsfxsize=3in \epsfbox{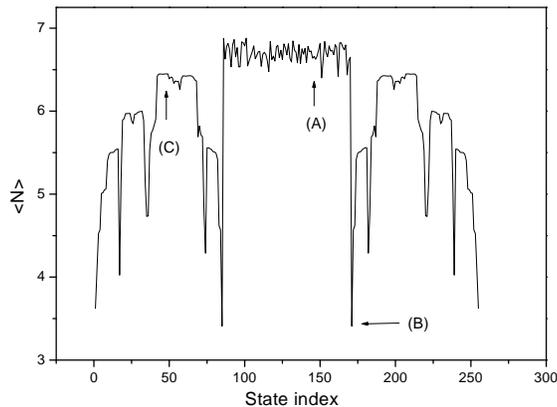}}
\caption{Total phonon number for each eigenstate.  Highest values
plateau for the most localized electronic amplitudes, A and C,
while lowest counts correspond to extended electronic states, such
as B.}
 \label{fig3}
\end{figure}

\begin{figure}[b]
\centerline{ \epsfxsize=4in \epsfbox{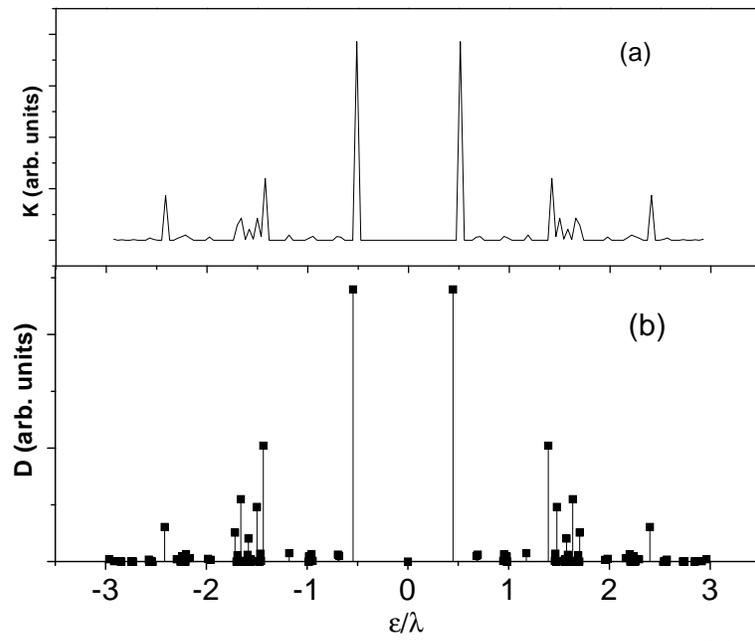}}
\caption{(a) Optical absorption spectrum $K$ within miniband
with level broadening added. (b)
Tunneling amplitude $D$.  Both cases for spectrum in Fig.\ 1.}
 \label{fig4}
\end{figure}


\end{document}